\documentclass[twocolumn,prl]{revtex4}
\usepackage{graphicx}
\usepackage{subfigure}
\usepackage{bm}
\usepackage{amssymb}
\newcommand{\Ket}[1]{\vert \, #1 \, \rangle}
\newcommand{\Bra}[1]{\langle \, #1 \,\vert}

\newcommand{\la}{\langle}
\newcommand{\ra}{\rangle}

\newcommand{\lp}{\left(}
\newcommand{\rp}{\right)}

\renewcommand{\phi}{\varphi}
\renewcommand{\epsilon}{\varepsilon}
\renewcommand{\vec}[1]{{\bf #1}}

\begin{document}

\title{Electrically-Driven Reverse Overhauser Pumping of Nuclear Spins in Quantum Dots}
\author{M. S. Rudner and L. S. Levitov}
\affiliation{
 Department of Physics,
 Massachusetts Institute of Technology, 77 Massachusetts Ave,
 Cambridge, MA 02139}

\date{\today}

\begin{abstract}
We propose a new mechanism for polarizing nuclear spins in quantum dots,
based on periodic modulation of the 
hyperfine coupling  by electric driving at
the electron spin resonance frequency. 
Dynamical nuclear polarization results from resonant excitation rather than 
hyperfine relaxation mediated by a thermal bath, and thus is
not subject to Overhauser-like detailed balance constraints.
This allows polarization in the direction
opposite to that expected from the Overhauser effect.
Competition of the electrically-driven and bath assisted mechanisms
can give rise to spatial modulation and sign reversal 
of polarization on a scale smaller than the electron confinement
radius in the dot. The relation to 
reverse Overhauser polarization observed in GaAs quantum dots
is discussed.
\end{abstract}

\maketitle

Advances in semiconductor quantum dot technology have given experimentalists unprecedented control over electron and nuclear spins, leading the way to many exciting applications in spintronics and quantum information processing.
Coherent control of electron spins has been demonstrated in few-electron double quantum dots \cite{Petta,DelftESR,HarvardESR}.
Because electron spin decoherence in GaAs quantum dots
arises mainly due to coupling to the disordered nuclear spin system \cite{KLG, CL}, there is great interest in developing new ways of controlling nuclear spins \cite{Imamoglu, Stepanenko}.

Nuclear spin polarization has been achieved in double dots
in the spin-blockaded DC transport regime 
where the hyperfine interaction couples the 
electron and nuclear spins \cite{Ono2004,Koppens2005,Baugh2007}.
Also, recent experiments on electron spin resonance (ESR)
in quantum dots\,\cite{DelftESR,HarvardESR} have demonstrated dynamical 
nuclear polarization
through resonant driving of ESR.
Interestingly, the polarization\,\cite{HarvardESR,Delft_private_comm} was found to be in the direction opposite
to that predicted by the thermodynamic arguments\,\cite{Overhauser,Abragam} 
that have successfully explained the direction of pumping in previous experiments \cite{OpticalOrientation, Gammon}.
Thus a new mechanism of nuclear spin pumping 
must be found.

\begin{figure}
  \flushleft \includegraphics[width=3.1in]{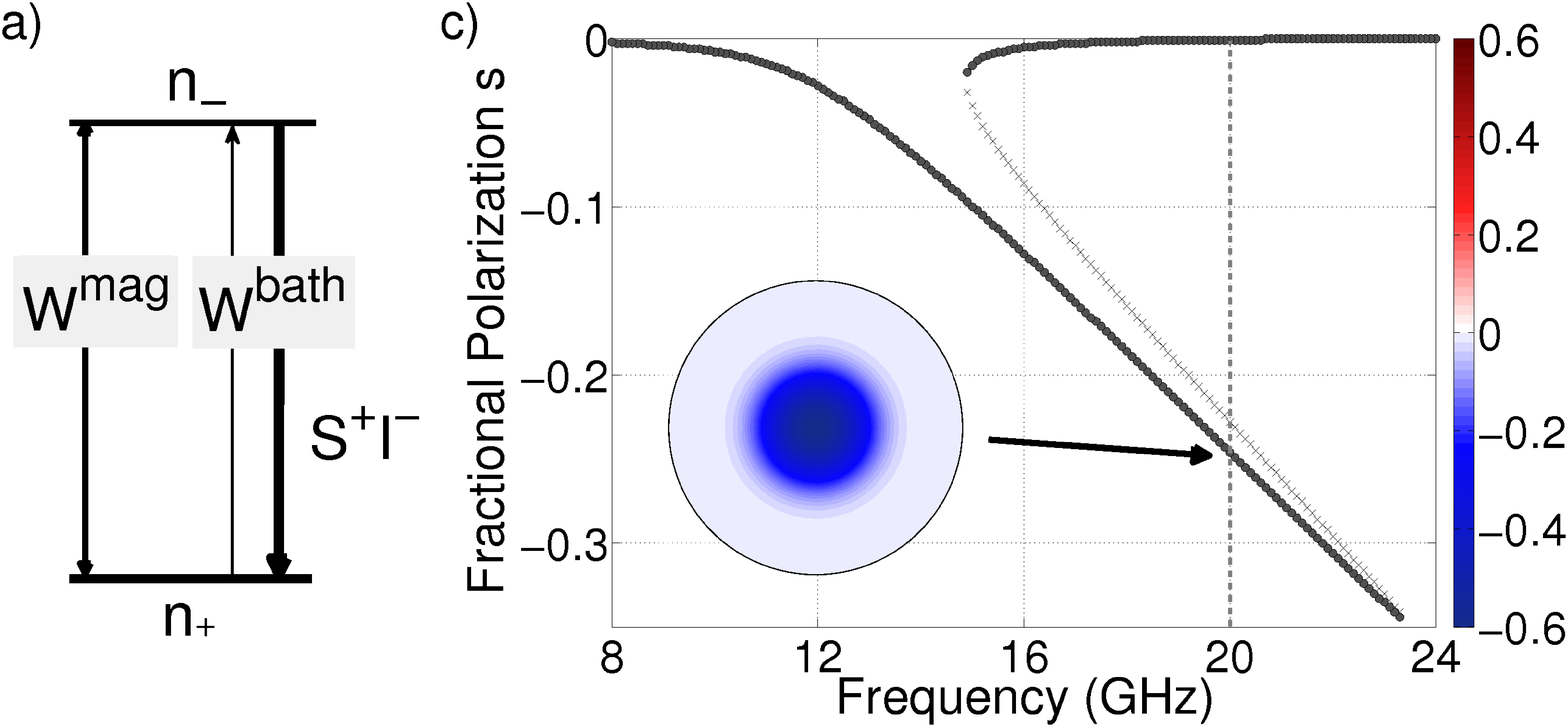}
  \flushleft \includegraphics[width=3.1in]{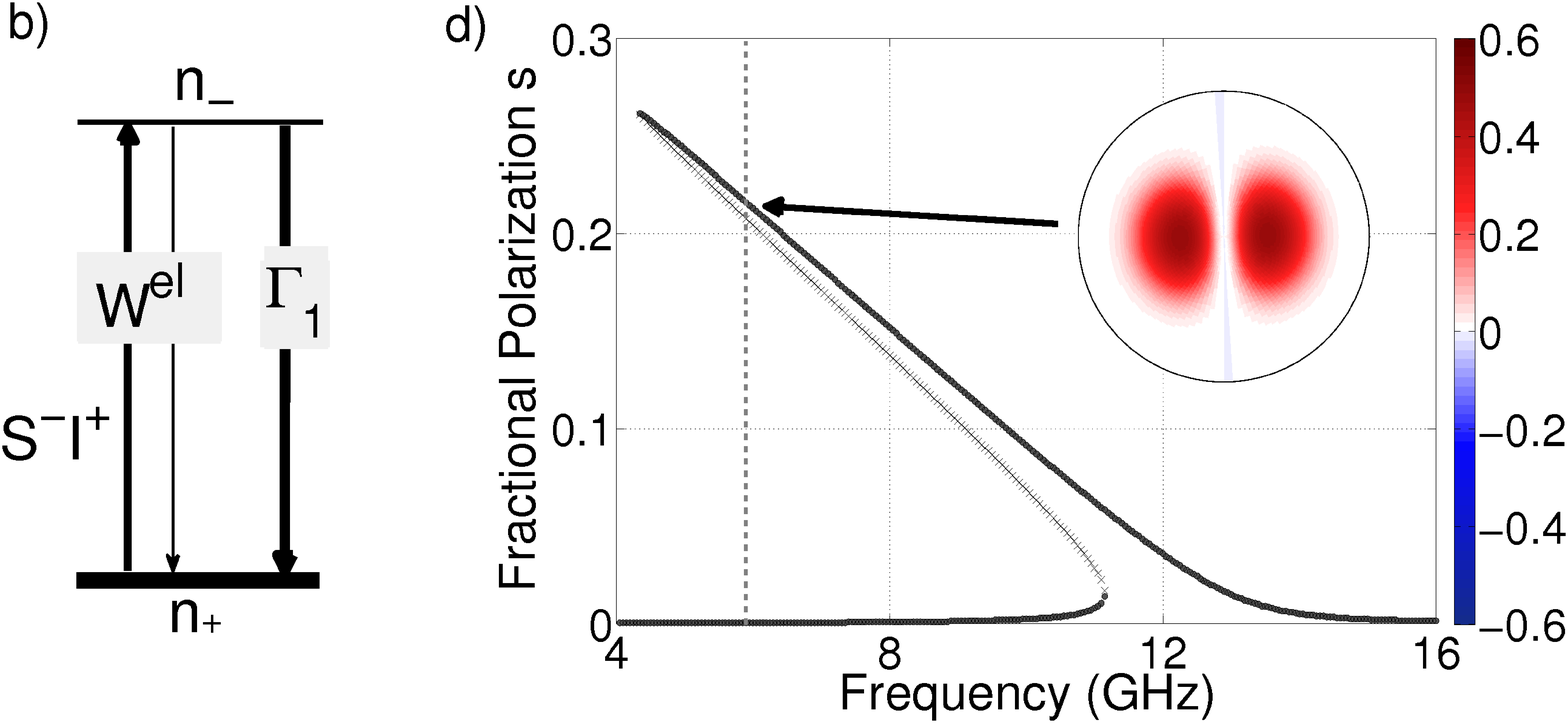}
\vspace{-3.5mm}
 \caption[]{
Nuclear polariation in a, c) the Overhauser and b, d) the reverse Overhauser regimes.
Panels a) and b) display the key transitions between electron Zeeman levels due to
magnetic/electric ESR driving, $W^{\rm mag,\, el}$, bath assisted hyperfine relaxation,
$W^{\rm bath}$, and electron spin relaxation, $\Gamma_1$.
Dominant electron-nuclear spin flip processes are marked by $S^{\pm} I^{\mp}$.
Panels c) and d) show the steady-state nuclear polarization determined self-consistently
via Eq.(\ref{Polarization}) combined with (\ref{LocalEquilibrium}) and (\ref{ESS}).
Insets show the spatial distribution of nuclear polarization in a circular region of
radius 50 nm, obtained using the electron density $|\psi(r)|^2 \propto e^{-r^2/r_0^2}$
with $r_0 = 25\, {\rm nm}$.
}
\label{StabilityDiagrams}
\end{figure}

The essence of the Overhauser pumping mechanism\,\cite{Overhauser,Abragam} is that a large mismatch of Zeeman energies prevents mutual electron-nuclear spin flips without coupling to an external bath, e.g. phonons.
Pumping occurs in the direction determined by the constraint that at low temperatures the bath can only accept energy from the system: spin flips occur primarily via spontaneous emission of excitations into the bath (see Fig.\ref{StabilityDiagrams}a).

Nuclear spin polarization is observed in \cite{DelftESR,HarvardESR} as an ESR frequency shift due to the longitudinal part of the hyperfine interaction between the electron spin ${\bm\hat{S}}$ and all nuclear spins $\{{\bm \hat{I}}_k\}$, 
   $ \hat{H}^{\rm HF} = A \sum_k \vert \psi(\vec{r}_k)\vert^2 \hat{{\bm S}}\cdot \bm{\hat{I}}_k $,
where $A$ is the hyperfine constant, $\psi(\vec{r})$ is the electron envelope wavefunction, and $\{\vec{r}_k\}$ are the positions of the nuclei.
Due to this shift, the electron Zeeman energy becomes:
\begin{equation}\label{eq:ESR_frequency}
\Delta\epsilon = | g\mu B_0 +  An_0s |,
\end{equation}
where $n_0$ and $s \sim \sum_k \la \hat{I}^z_k\ra$ are the density and effective  
polarization of nuclei (see Eq.(\ref{Polarization})).
The direction of the frequency shift in response to nuclear polarization is determined by the relative signs of $A$, $\mu$, and the electron $g$-factor.
Because these quantities are all negative in GaAs 
($g \approx -0.4$, $An_0 \approx -130\, \mu {\rm eV}$), 
the thermodynamic argument\,\cite{Overhauser,Abragam} predicts a positive
frequency shift in contrast to the observed {\it negative} frequency shift \cite{HarvardESR,Delft_private_comm}.

In this work we consider the situation 
where hyperfine flip-flop transitions are stimulated 
by an externally applied time-varying electric field as in \cite{HarvardESR}.
In this case, the direction of pumping is not set by the relaxational energy balance of hyperfine transitions.
Rather, pumping occurs when other mechanisms of electron spin relaxation replenish the population of the electron spin ground state without coupling to nuclear spins (see Fig.\ref{StabilityDiagrams}b), thus allowing the electron to repeatedly pump many nuclear spins in the direction selected by electron spin excitation.

We show below that this mechanism leads to ``reverse Overhauser'' pumping.
To compare the Overhauser and reverse Overhauser mechanisms, we add magnetic driving and bath-mediated hyperfine relaxation of nuclear spins to the model.
By solving a self-consistency equation for the effective nuclear polarization $s$, we map out the fixed points of polarization
(see Fig.\ref{StabilityDiagrams}). 
An interesting spatial modulation of nuclear polarization
on a scale less than the electron confinement radius
is predicted in the electrically driven regime.
In the crossover regime where these mechanisms act in direct competition, 
we find complex, history-dependent ESR shifts.

We note that ``inverted'' Overhauser pumping has been observed 
previously in other systems\,\cite{Bennet}.
The thermodynamic basis for that pumping was Overhauser-like; the direction of pumping was reversed because spin flips were caused by non-secular terms of the dipole-dipole interaction such as $\hat{S}^+ \hat{I}_k^+$\,\cite{Abragam}.
However, because the electron-nuclear dipole coupling is weak in GaAs this mechanism  cannot explain the results of experiments \cite{DelftESR} and \cite{HarvardESR}.

The driving of ESR by coupling a localized electron to an electric field ${\rm \bf E}(t)\propto {\rm \bf E}_\omega e^{-i\omega t}$ in the presence of an inhomogeneous nuclear spin distribution is described, in the harmonic approximation \cite{HarvardESR}, by the Hamiltonian: 
\begin{eqnarray}
  \label{elDriveHam} \hat{H}^{\rm el}(t) = \frac{A}{2}\sum_k \vec{d}_\omega \cdot \nabla \vert \psi(\vec{r}_k)\vert^2 \lp \hat{S}^+ \hat{I}_k^{-} e^{-i \omega t} + {\rm h.c.}\rp,
\end{eqnarray}
where $\vec{d}_\omega \propto {\vec E}_\omega$ is the electron displacement due to the AC
field \cite{HarvardESR}.
Because the hyperfine interaction conserves spin, electron spin flips caused by (\ref{elDriveHam}) are accompanied by compensating nuclear spin flips in the opposite direction.

We assume that electron decoheres quickly enough to prevent the entanglement of electron and nuclear spins so that we can describe the system by the factorized density matrix $\hat{\rho} = \hat{\rho}_{\rm el} \otimes \hat{\rho}_{\rm N}$.
The time for this entanglement to build up was analyzed in Ref.\cite{YLS} and found to be: $T_{\rm ent} \approx N \Delta\epsilon/(An_0)^2$, where $N$ is the number of nuclear spins in the system. 
If the electron decoherence rate $\Gamma_2 > 1/T_{\rm ent}$, then this entanglement is inessential for our analysis.

Furthermore, we assume that the electron spin decoheres much faster than the rate of spin flips induced by (\ref{elDriveHam}), which allows these spin flip processes to be treated as incoherent.
In this limit, described by $\hat{\rho}_{\rm el}=n_+|\uparrow\ra\la\uparrow|+n_-|\downarrow\ra\la\downarrow|$,
the transition rate can be calculated using Fermi's Golden Rule.

Our main motivation for working in the incoherent regime
is that it makes the physics of the reverse Overhauser mechanism most apparent;
in this regime it is straightforward to understand and exhibit the principle, and to obtain the
sign of the effect.
Additionally, it may provide a good framework to understand experiment \cite{HarvardESR}
where coherent Rabi oscillations were not observed. 
While coherence in some systems (e.g. \cite{DelftESR}) 
may be higher than assumed in our model, 
we do not expect that it can reverse the direction of dynamical polarization.

For simplicity we consider the case of spin-1/2 nuclei and calculate the spin flip rates for each nucleus independently using a nuclear state of the form 
\[
\hat{\rho}_N = \bigotimes_k\,\lp N_{k, +} \Ket{\uparrow}\Bra{\uparrow} + N_{k, -} \Ket{\downarrow}\Bra{\downarrow}\,\rp.
\]
Here $N_{k, \pm}$ is the occupation probability of the up (down) spin state 
of nucleus $k$, with $N_{k, +} + N_{k, -} = 1$.

After a straightforward calculation, we find the spin flip transition rate $W^{\rm el}_k$ for the nucleus at position $\vec r_k$
\begin{equation}
  \label{elRate} W_{k}^{\rm el} = \frac{A^2}{4} \left(\vec{d}_\omega \cdot \nabla |\psi(\vec r_k)|^2\right)^2 \frac{\Gamma_2}{(\omega-\Delta\epsilon)^2+(\Gamma_2/2)^2},
\end{equation}
where $\Delta\epsilon$ is the electron Zeeman energy (\ref{eq:ESR_frequency}), measured in units of frequency.
Parenthetically, because the electrical excitation involves simultaneous electron and nuclear spin flips, the energy $\Delta\epsilon$ should include the nuclear Zeeman energy which is small and will be ignored hereafter.
The effective fractional polarization of nuclei contributing to the Overhauser shift is
\begin{eqnarray}
  \label{Polarization} s \equiv \sum_k|\psi(\vec r_k)|^2 \delta V (N_{k,+}-N_{k,-})
,
\end{eqnarray}
where $\delta V$ is the unit cell volume, $\delta V=n_0^{-1}$.

The net spin-flip rate for nucleus $k$ is determined by the upward and downward flip rates due to electric driving with rate $W^{\rm el}_k$ and out-diffusion with rate $\gamma$:
\begin{eqnarray}
  \label{FlipRates} 
  \dot{N}_{k, +} = W^{\rm el}_k(n_+N_{k,-} - n_-N_{k,+}) + \gamma (N_{k,-} - N_{k,+}).
\end{eqnarray}
Similarly, the electron state evolves according to
\begin{eqnarray}
  \label{elFlipRate} \dot{n}_+ =\sum_kW^{\rm el}_k(n_-N_{k,+} - n_+N_{k,-}) + \Gamma_1 n_- - \tilde{\Gamma}_1n_+.
\end{eqnarray}
The electron spin flip rate includes contributions from all nuclei.
The forward and reverse relaxation rates are related by a Boltzmann factor $\tilde{\Gamma}_1 = e^{-\beta \Delta\epsilon} \Gamma_1$ in accordance with detailed balance.
The incoherent driving model (\ref{FlipRates}), (\ref{elFlipRate}) is valid when $\Gamma_2 \gg W^{\rm el} \equiv \sum_k W^{\rm el}_k$.

The resulting polarization is determined by the steady states of the combined system (\ref{FlipRates}) and (\ref{elFlipRate}).
The sign of polarization can be exhibited immediately by summing Eq.(\ref{FlipRates}) over all $k$ and combining it with Eq.(\ref{elFlipRate}):
\begin{eqnarray}
  \label{LLCondition} \sum_k (N_{k,+} - N_{k,-}) = \frac{\Gamma_1}{\gamma}\left(n_- - e^{-\beta \Delta\epsilon}n_+\right).
\end{eqnarray}
ESR driving upsets the electron equilibrium, which makes $n_- \ge e^{-\beta \Delta\epsilon} n_+$ and leads to positive (reverse Overhauser) polarization (\ref{LLCondition}). 
Even at weak driving when $n_- \approx e^{-\beta \Delta\epsilon} n_+$, the polarization can still be large if nuclear relaxation is sufficiently slow: $\gamma \ll \Gamma_1$.

To compare our mechanism with the Overhauser pumping mechanism, and to facilitate the discussion of steady states, it is useful to introduce driving due to a transverse AC magnetic field of strength $B_1$
\begin{eqnarray}
  \label{magRate} W^{\rm mag} = \frac{1}{4}\frac{(g \mu_B B_1)^2 \Gamma_2}{(\omega - \Delta\epsilon)^2 + (\Gamma_2/2)^2},
\end{eqnarray}
which adds an additional term  $W^{\rm mag}(n_- - n_+)$ to (\ref{elFlipRate}).
As before, we assume $W^{\rm mag} \ll \Gamma_2$.

In principle, $\Gamma_1$ and $\Gamma_2$ can differ by many orders of magnitude.
When $\Gamma_1 \ll \Gamma_2$, our rate equation approach can treat both the weak driving $W^{\rm mag,\, el} \ll \Gamma_1$ and strong driving $\Gamma_1 < W^{\rm mag,\, el} \ll \Gamma_2$ limits.
If $\Gamma_1 \sim \Gamma_2$, however, the rate equation approach is valid in the weak driving limit. 
To treat strong driving in this regime, one must solve the full Bloch equations for electron spin \cite{Danon}.

The Overhauser mechanism involves bath-mediated hyperfine relaxation of nuclear spins. 
This rate includes a product of two factors describing the coupling to the bath and the hyperfine interaction, giving spatial dependence $W^{\rm bath}_k \propto\vert\psi(r_k)\vert^4$. 
To account for this process we add the term $W^{\rm bath}_k(N_{k,-} n_+ - e^{-\beta \Delta\epsilon} N_{k,+} n_-)$ to (\ref{FlipRates}).

The local steady state of nuclear spins is descibed by
\begin{eqnarray}
  \label{LocalEquilibrium} \frac{N_{k,+}}{N_{k,-}} = \frac{(W^{\rm el}_k+\tilde W^{\rm bath}_k)n_+ + \gamma}{(W^{\rm el}_k + W^{\rm bath}_k)n_- + \gamma}.
\end{eqnarray}
The competition between the Overhauser and reverse Overhauser mechanisms exhibited by Eq.(\ref{LocalEquilibrium}) is simplest to understand in the absence of nuclear spin relaxation, $\gamma = 0$.
When electric driving is weak compared with bath-mediated relaxation, $W^{\rm el}_k < W^{\rm bath}_k$, saturating ESR by $W^{\rm mag}$ makes $n_+ \approx n_-$ and causes the nuclei to polarize opposite to the external field: $N_{k,+}/N_{k,-} \approx \exp(-\Delta\varepsilon/k_B T) < 1$.
However, if $W_k^{\rm bath}$ is small compared with $W_k^{\rm el}$, then Eq.(\ref{LocalEquilibrium}) simplifies to $N_{k,+}/N_{k,-} \approx n_+/n_-$.
Due to electron relaxation, $n_+/n_- > 1$ and nuclei polarize in the reverse Overhauser direction.

To describe this competition quantitatively, we consider a simple limit $W^{\rm mag} \gg W^{\rm el}$ where
\begin{eqnarray}
  \label{ESS} \frac{n_+}{n_-} \approx \frac{W^{\rm mag} + \Gamma_1}{W^{\rm mag} + \tilde{\Gamma}_1}.
\end{eqnarray}
Although electric driving has a negligible effect on the electron state in this limit, it still dominates the nuclear dynamics when $W^{\rm bath}_k < W^{\rm el}_k$.
If electric driving is not weak, the rates $\sum_k W^{\rm el}_k N_{k,\pm}$ appear in the numerator and in the denominator of Eq.(\ref{ESS}).
This complicates the mathematics without changing the results qualitatively. 

\begin{figure}
\includegraphics[width=3.1in]{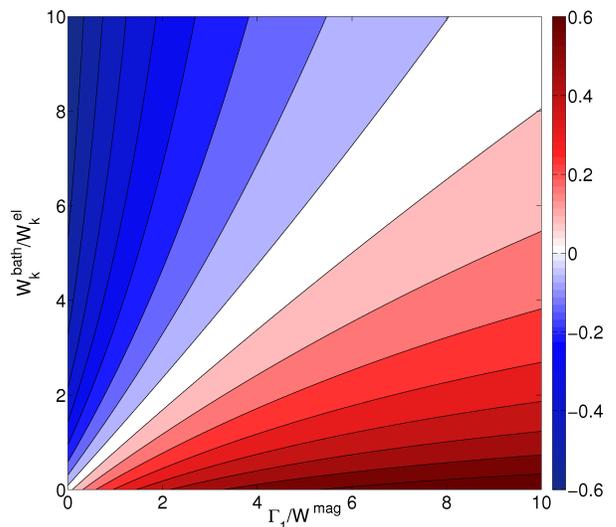}
\vspace{-3.5mm}
 \caption[]{Local steady-state nuclear polarization $s_k \equiv N_{k,+}- N_{k,-}$ versus $W^{\rm bath}_k/W^{\rm el}_k$ and $\Gamma_1/W^{\rm mag}$.
We obtain $s_k$ from (\ref{LocalEquilibrium}) with $\beta\Delta\epsilon = 2$ (e.g. $B_0 = 750\, {\rm mT}$ and $T = 100\, {\rm mK}$) and $\gamma = 0$.
Reverse Overhauser pumping dominates below the main diagonal.
Due to the spatial dependence of $W^{\rm bath}_k/W^{\rm el}_k$, polarizations in different regions of the dot are described by different points in this diagram (see Figs. \ref{StabilityDiagrams} and \ref{Crossover}).}
\label{fig:Polarization}
\end{figure}

The competition between the two pumping mechanisms, controlled by the relative strengths of $W_k^{\rm el}$ and $W_k^{\rm bath}$ and of $W^{\rm mag}$ and $\Gamma_1$ via (\ref{LocalEquilibrium}) and (\ref{ESS}), is illustrated in Fig.\ref{fig:Polarization}. 
As expected, the reverse Overhauser effect is strongest when $\Gamma_1/W^{\rm mag} \gg 1$ and $W_k^{\rm el}/W_k^{\rm bath} \gg 1$, i.e. when electron spin relaxation and electrically-driven spin-flip transitions are strong compared to magnetic ESR driving and 
hyperfine relaxation, respectively.

So far we have considered the local polarization described by Eq.(\ref{LocalEquilibrium}) and Fig.\ref{fig:Polarization}.
These results can be directly applied to the entire system in the limit where internal diffusion of nuclear spins homogenizes the polarization distribution over a time scale much shorter than the time scale for polarization build-up. 
Because nuclear spin diffusion can be slow \cite{KLG}, we must consider the full spatial dependence of polarization.

To understand the nuclear steady states, 
it is necessary to account for the resonant character of the rates (\ref{elRate}) and (\ref{magRate}), which are sensitive to the total polarization $s$ due to the hyperfine shift that brings the system in-to or out-of resonance.
To this end, we determine the steady-state values of polarization self-consistently by combining expressions (\ref{LocalEquilibrium}) and (\ref{ESS}) with the definition of $s$, Eq.(\ref{Polarization}).
This gives a self-consistency condition of the form $s = f(s)$, where the function $f(s)$ is peaked near the value of $s$ where the Overhauser shift brings the electron Zeeman energy into resonance with the driving field.
Depending on parameter values, one or more solutions may exist.

The stable and unstable polarization fixed points 
are plotted in Fig.\ref{StabilityDiagrams} as a function of driving frequency for parameters deep in the Overhauser and reverse Overhauser regimes
with $B_0 = 2\, {\rm T}$, $\Gamma_1 = \Gamma_2 = 10^8\, s^{-1}$, and $\gamma = 0.05\ s^{-1}$.
In the Overhauser regime we use
$B_1 = 5\, {\rm mT}$ and $W^{\rm bath} = 5\, s^{-1}$, 
while in the reverse Overhauser regime we use
$B_1 = 1\, {\rm mT}$ and  $d_\omega = 1\, {\rm nm}$.
These parameter values were chosen, in a realistic range \cite{DelftESR, HarvardESR, Imamoglu, EN}, to clearly exhibit the behavior in the two pumping regimes.
When comparing the values of $\Gamma_{1,2}$ with those expected in quantum dots, the reader should bear in mind that these rates are highly sensitive to coupling to the leads \cite{Dreiser} and driving strength.


When the Overhauser mechanism dominates, nuclear spins polarize in the direction that adds to the external field and 
``resonance-dragging'' is observed
as frequency is swept from low to high.
If the reverse Overhauser mechanism dominates,  
nuclei polarize in the opposite direction and the system remains on resonance 
on the sweep from high to low frequencies.
Analogous hysteretic behavior occurs when sweeping magnetic field.

Spatial dependence of polarization
arises due to the spatial variations of the rates $W_k^{\rm bath} \propto \vert\psi(r_k)\vert^4$ and $W_k^{\rm el} \propto \left\vert \nabla_x \vert \psi(r_k)\vert^2\right\vert^2$, where 
without loss of generality we have taken the external driving $\vec E(t)\parallel \hat x$.
Once the steady state values of the net polarization $s$ 
are identified, the local nuclear polarization at each point $r_k$ can be found by substituting 
$s$
into Eq.(\ref{LocalEquilibrium}).
The resulting polarization distributions $s(r_k)=N_{k,+} - N_{k,-}$,
obtained for a Gaussian electron wavefuntion $|\psi(r)|^2 \propto e^{-r^2/r_0^2}$, are displayed in the insets of Fig.\ref{StabilityDiagrams}. 

Due to the dependence of $W_k^{\rm el}$ on the gradient of the wave function, 
nuclei on the shoulders of $|\psi(x)|^2$ are 
most strongly affected by the reverse Overhauser mechanism, 
while nuclei near the center of the dot are not affected at all. 
In the reverse Overhauser regime, polarization builds up in lobes where the derivative 
$\nabla_x|\psi|^2$ is large (Fig.\ref{StabilityDiagrams}b).
Conversely, nuclear spins near the center of the dot where $|\psi(r)|^2$
is maximal are most strongly pumped by the Overhauser mechanism (Fig.\ref{StabilityDiagrams}a).

\begin{figure}
\includegraphics[width=3.1in]{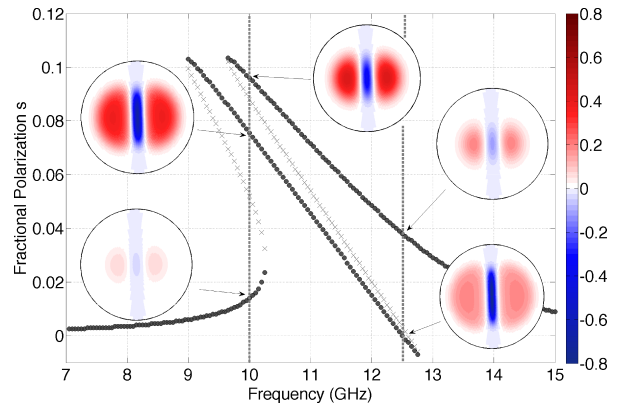}
\vspace{-3.5mm}
 \caption[]{Stability diagram in the crossover regime with $B_0 = 2\, {\rm T}, B_1 = 2.5\, {\rm mT}, d_\omega = 1\, {\rm nm}, \Gamma_1 = \Gamma_2 = 10^8\, s^{-1}, W^{\rm bath} = 1\, s^{-1}$, $\gamma =0.005\, s^{-1}$.
Insets show nuclear polarization distributions in the stable states indicated by the arrows.
ESR frequency dragging 
occurs in both sweep directions.
The effect is asymmetrical, with maximum shift of 300 MHz in the positive direction and 3.5 GHz in the negative direction.
}
\label{Crossover}
\end{figure}

A more complex behavior is featured by the crossover regime where both mechanisms are of comparable strength.
In this case there are additional stable states of nuclear polarization,
organized as shown in Fig.\ref{Crossover}.
The corresponding spatial distributions of polarization are 
indicated by the arrows.
While the net polarization remains not too large, 
the system exhibits separate regions of very large positive and negative polarization.
Interestingly, we find that there exists a region of parameter values for which ``resonance-dragging'' in the crossover regime
can occur in {\it both} frequency sweep directions.

 Reverse Overhauser pumping 
 was observed when ESR was driven both 
 electrically\,\cite{HarvardESR} and magnetically\,\cite{DelftESR,Delft_private_comm}.
 Although superficially this might seem to present a problem, 
 it is not inconceivable that a stray microwave electric field 
was present in experiment\,\cite{DelftESR}, 
which was too weak to significantly contribute to driving ESR 
but strong enough to alter the nuclear polarization pumping mechanism.
This hypothesis could be tested, e.g. by applying an oscillating voltage 
to the side gate along with the strip-line microwaves.
By adjusting the relative phase between this perturbation and the microwaves it should be possible to enhance or partially cancel the field responsible for electrically-induced transitions, resulting in a reversal of
the polarization pumping direction.

Experimentally, the clearest evidence in favor of the proposed mechanism
could be obtained,
as suggested by Fig.\ref{fig:Polarization}, 
by independently varying the electron spin decay rates
$\Gamma_1$ and $W^{\rm bath}$ and the ESR excitation strengths $W^{\rm mag}$ and $W^{\rm el}$
in order to observe polarization reversal.
Although the mechanism of electron spin relaxation 
is uncertain, a likely candidate is cotunneling to the leads 
resulting in spin exchange\,\cite{Delft_private_comm, Dreiser}. The same may also 
apply for the hyperfine spin flip rate $W^{\rm bath}$,
which would make both $\Gamma_1$ and $W^{\rm bath}$ sensitive 
to the strength of coupling to the leads that can be adjusted {\it in situ}.

We also comment on an alternative explanation of reverse Overhauser
pumping \cite{HarvardESR} that assumes unequal loading probability
of the Zeeman-split electron spin 
states. 
These mechanisms can be distinguished by studying pumping efficiency
as a function of the time each electron is held in the dot: the mechanism based on unequal loading probability predicts a
polarization buildup rate proportional to the rate at which the electron state
is refreshed while the mechanism discussed above remains
efficient even if a single electron is kept in the dot for a long time.


Besides offering an explanation of the anomalous sign of 
polarization \cite{HarvardESR}, the proposed mechanism of nuclear pumping by electrically driven ESR provides a new tool for controlling nuclear spins.
Combined with conventional Overhauser pumping, it allows 
polarized nuclear states of either sign with spatial modulation
on a scale less than the electron confinement radius
to be created.

We benefited from useful discussions with C. Barthel, D. G. Cory, J. Danon, F. H. L. Koppens, E. A. Laird, C. M. Marcus, Yu. V. Nazarov, E. I. Rashba, and L. M. K. Vandersypen,
and partial support from W. M. Keck foundation.  
M. R.'s work was supported by DOE CSGF, Grant No. DE-FG02-97ER25308.


\end{document}